\documentclass[sigconf]{acmart}
\usepackage{booktabs} 
\usepackage[flushleft]{threeparttable}
\usepackage{multirow}
\usepackage{bbm}
\usepackage{subcaption}
\usepackage{bm}
\usepackage{physics}
\usepackage{balance}
\usepackage{amsmath}
\usepackage[linesnumbered,boxed,ruled,vlined]{algorithm2e}
\newcommand{\eat}[1]{}

\copyrightyear{2020}
\acmYear{2020}
\setcopyright{iw3c2w3}
\acmConference[WWW '20]{Proceedings of The Web Conference 2020}{April 20--24, 2020}{Taipei, Taiwan}
\acmBooktitle{Proceedings of The Web Conference 2020 (WWW '20), April 20--24, 2020, Taipei, Taiwan}
\acmPrice{}
\acmDOI{10.1145/3366423.3380171}
\acmISBN{978-1-4503-7023-3/20/04}

\begin{document}
\title{Adversarial Attack on Community Detection by Hiding Individuals}


\author{Jia Li}
\affiliation{%
  \institution{The Chinese University of Hong Kong}
}
\email{lijia@se.cuhk.edu.hk}

\author{Honglei Zhang}
\affiliation{%
  \institution{Georgia Institute of Technology}
}
\email{zhanghonglei@gatech.edu}

\author{Zhichao Han}
\affiliation{%
  \institution{The Chinese University of Hong Kong}
}
\email{zchan@se.cuhk.edu.hk}

\author{Yu Rong}
\affiliation{%
  \institution{Tencent AI Lab}
}
\email{yu.rong@hotmail.com}

\author{Hong Cheng}
\affiliation{%
  \institution{The Chinese University of Hong Kong}
}
\email{hcheng@se.cuhk.edu.hk}

\author{Junzhou Huang}
\affiliation{%
  \institution{Tencent AI Lab}
}
\email{joehhuang@tencent.com}

\renewcommand{\shortauthors}{J. Li et al.}

\begin{abstract}
It has been demonstrated that adversarial graphs, i.e., graphs with imperceptible perturbations added, can cause deep graph models to fail on node/graph classification tasks.  In this paper, we extend adversarial graphs to the problem of community detection which is much more difficult.  We focus on black-box attack and aim to hide targeted individuals from the detection of deep graph community detection models, which has many applications in real-world scenarios, for example, protecting personal privacy in social networks and understanding camouflage patterns in transaction networks.  We propose an iterative learning framework that takes turns to update two modules: one working as the constrained graph generator and the other as the surrogate community detection model.  We also find that the adversarial graphs generated by our method can be transferred to other learning based community detection models. \eat{Together with this work, we publish a large community detection dataset which consists of 528,672 nodes and 1,149 non-overlapping communities.}

\end{abstract}

%
%




\maketitle

\section{Introduction}


Community detection is one of the most widely studied topics in the graph domain, which aims to discover groups of nodes in a graph, such that the intra-group connections are denser than the inter-group ones \cite{wang2015community}.  It has been widely applied to many real-world applications ranging from functional module identifications in a protein-protein interaction network \cite{ahn2010link}, scientific discipline discoveries in a coauthor network \cite{zhou2009graph}, to fraud organization detections in a user-user transaction network \cite{akoglu2015graph}. However, with the rapid development of the community detection algorithms, people realize that their privacy is over-mined \cite{Chen2018GABasedQO}. In this context, some work begins to investigate the techniques that allow to hide individuals, communities \cite{waniek2018hiding,fionda2017community} or degrade the overall performance of community detection algorithms \cite{Chen2018GABasedQO}, mostly based on heuristics or genetic algorithms.

Recently deep graph learning models \cite{perozzi2014deepwalk,kipf2017semi} have achieved remarkable performance in many graph learning tasks.  Meanwhile, some studies \cite{zugner2018adversarial,dai2018adversarial} notice that deep graph models can be easily attacked on tasks such as node/graph classification.  Motivated by these findings, in this work, we aim to address the following question: \emph{how vulnerable are graph learning based community detection methods? Can we hide individuals by imperceptible graph perturbations?}  A good solution to this problem can benefit many real-world applications, e.g., personal privacy protection, fraud escape understanding.

\begin{figure*}
\begin{center}
\includegraphics [width=0.95\textwidth]{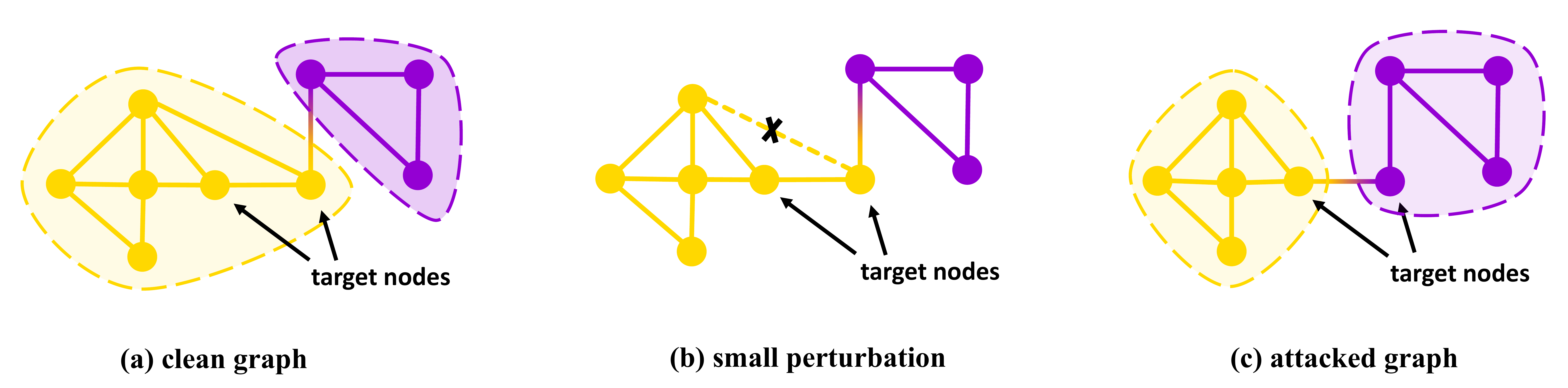}
\end{center}

\caption{An example of adversarial attack on community detection by hiding individuals.  At first, the two targets are clustered into one community (Fig.\ 1(a)).  With a small perturbation of deleting an edge (Fig.\ 1(b)), the two targets are assigned into different communities by the same detection method (Fig.\ 1(c)).}
\label{fig.example1}
\end{figure*}

Unlike adversarial attacks on node/graph classification where gradients \cite{zugner2018adversarial} or limited binary responses \cite{dai2018adversarial} of the target classifier are available, one challenge we face is that there is no feedback from the target model. For example, in social networking companies like Facebook or Twitter, community detection algorithms are serving as a backend for other purposes such as advertising, which prevents the direct interactions between the target model and individuals. To tackle this challenge, we design a surrogate community detection model which is based on the widely used graph neural networks (GNNs) \cite{kipf2017semi} and a popular community detection measure \emph{normalized cut} \cite{shi2000normalized}.  We attack this surrogate community detection model and verify that the attack can also be transferred to other popular graph learning based community detection models.

In the literature of adversarial attack on graph data, one commonly observed difficulty is that it is very hard to quantify the adversarial costs of various attacks \cite{sun2018adversarial}.  While imperceptible perturbations can be checked by human in the image domain, it is impossible to adopt the same strategy in the graph domain.  Currently, most existing methods tackle this indirectly with either a discrete budget to limit the number of allowed changes \cite{sun2018adversarial, dai2018adversarial} or a predefined distribution such as power-law distribution \cite{zugner2018adversarial}.  However, the former based on discrete changes is helpful but far from sufficient, whereas the latter emphasizing a power-law distribution is proven to be rare in reality \cite{broido2019scale}.  In this work, we propose a clearly comprehensible graph objective to measure the degree of perturbations from two perspectives: local proximity and global proximity.

Another challenge we face is the huge computation space of selecting proper candidate edges/nodes to modify.  It is non-trivial to develop a solution that can scale with the size of graphs.  Existing solutions such as \cite{waniek2018hiding} rely on heuristics to bypass this problem, which, however, fail to derive optimal choices, especially for attributed graphs.  In this work, we design a novel graph generation model which learns to select the proper candidates. With this approach, we can not only generate proper adversarial graphs to attack community detection models, but also explicitly take the imperceptible perturbation requirement into the learning process.

Our contributions are summarized as follows.

\begin{itemize}
\item We study adversarial attack on graph learning based community detection models via hiding a set of nodes, which, to the best of our knowledge, has not been studied before.  Our proposed solution CD-ATTACK achieves superior attack performance to all competitors.

\item We propose a new graph learning based community detection model which relies on the widely used GNNs and the popular measure \emph{normalized cut}. It serves as the surrogate model to attack; furthermore, it can also be used for solving general unsupervised non-overlapping community detection problems.

\item We define a comprehensible graph-related objective to measure the adversarial costs for various attacks from two perspectives: local proximity and global proximity.

\item We design a novel graph generation neural network that can not only produce adversarial graphs to community detection algorithms, but also satisfy the discrete constraint.

\item We evaluate CD-ATTACK on four real-world data sets. Our method outperforms competing methods by a large margin in two measures. In addition, we validate that the adversarial graphs generated by our method can be transferred to two other popular graph learning based community detection models.

\end{itemize}

The remainder of this paper is organized as follows.  Section \ref{def} gives the problem definition and Section \ref{alt} describes the design of CD-ATTACK.  We report the experimental results in Section \ref{sec.exp} and discuss related work in Section \ref{sec.related}.  Finally, Section \ref{sec.con} concludes the paper.

\section{Problem Definition}\label{def}
We denote a set of nodes as $V=\{v_1, v_2, \ldots, v_N\}$ which represent real-world entities, e.g., authors in a coauthor network, users in a user-user transaction network.  We use an $N\times N$ adjacency matrix $A$ to describe the connections between nodes in $V$.  $A_{ij}\in \{0, 1\}$ represents whether there is an undirected edge between nodes $v_i$ and $v_j$ or not, e.g., a coauthored paper that links two authors, a transaction that connects two users.  In this study, we focus on an undirected graph; yet our methodology is also applicable to directed graphs.  We use $X=\{x_1, x_2, \ldots, x_N\}$ to denote the attribute values of nodes in $V$, where $x_i\in \mathbb{R}^{d}$ is a $d$-dimensional vector.

The community detection problem aims to partition a graph $G = (V, A, X)$ into $K$ disjoint subgraphs $G_i = (V_i, A_i, X_i)$, $i=1, \ldots, K$, where $V = \cup_{i=1}^KV_i$ and $V_i\cap V_j = \emptyset$ for $i\neq j$.  In our study, we adopt this formulation which produces non-overlapping communities.

In the context of community detection, we are interested in a set of individuals $C^+ \subseteq V$ who actively want to escape detection as a community or part of a community, e.g., users who are involved in some underground transactions and thus eager to hide their identity, or malicious users who intend to fool a risk management system.  Given a community detection algorithm $f$, the \emph{community detection adversarial attack} problem is defined as learning an attacker function $g$ to perform small perturbations on $G = (V, A, X)$, leading to $\hat{G} = (\hat{V},\hat{A},\hat{X})$, such that

\begin{align}
\begin{split}
  \max{}& \mathcal{L}(f(\hat{G}), C^+) - \mathcal{L}(f(G), C^+)\\
  \text{s.t.}&\ \ \ \hat{G} \leftarrow \arg\min g(f, (G, C^+))\\
  &\ \ \ Q(G, \hat{G}) < \epsilon,
\end{split}
\end{align}
where $\mathcal{L}(\cdot, \cdot)$ measures the quality of community detection results with respect to the target $C^+$, $Q(G, \hat{G}) < \epsilon$ is used to ensure imperceptible perturbations.  In this work, we focus on edge perturbations such as edge insertion and deletion, i.e., $\hat{G} = (\hat{V}, \hat{A}, \hat{X}) = (V, \hat{A}, X)$.  Intuitively, we want to maximize the decrease of the community detection performance related to a subset $C^+$ by injecting small perturbations.

Figure \ref{fig.example1} depicts an example of community detection adversarial attack in a user-user transaction network.  At the beginning, the two target individuals are clustered into a community in yellow, which corresponds to a money laundering group.  One target, as a member of the community, would be suspected of money laundering given the other is exposed somehow.  With a small perturbation by deleting an edge, the target individuals are assigned into two communities, where the community in purple is a high-credit user group.  Thus the two targets decrease the probability of being detected as a whole.

\section{Methodology}\label{alt}
\subsection{Framework}
In our problem setting, we have two main modules: (1) an adversarial attacker $g(\cdot)$ that aims to perform \emph{unnoticeable} perturbations to the original graph such that a set of individuals could be hidden, and (2) a community detection algorithm $f(\cdot)$ that can partition a graph into several subgraphs in an unsupervised way.  These two modules are highly interrelated, i.e., the adversarial attacker $g(\cdot)$ needs the feedback from $f(\cdot)$ to check if the goal of hiding is achieved or not, and the community detection algorithm $f(\cdot)$ relies on adversarial examples to enhance its robustness.  In the literature, most studies achieve this interaction by exploiting the gradient or other moments of a differentiable loss function \cite{zugner2018adversarial}, which, however, is intractable in discrete data such as graph.  To bypass this difficulty, motivated by \cite{konda2000actor}, we utilize policy gradient in the Actor-Critic framework as the signal between the interaction of the two modules.  Another point is that, as we focus on black-box attack meaning there is no specific community detection algorithm in hand, we need to instantiate ourselves with a surrogate model with the capacity of generalization and robustness.

\begin{figure}
\begin{center}
\includegraphics [width=0.46\textwidth]{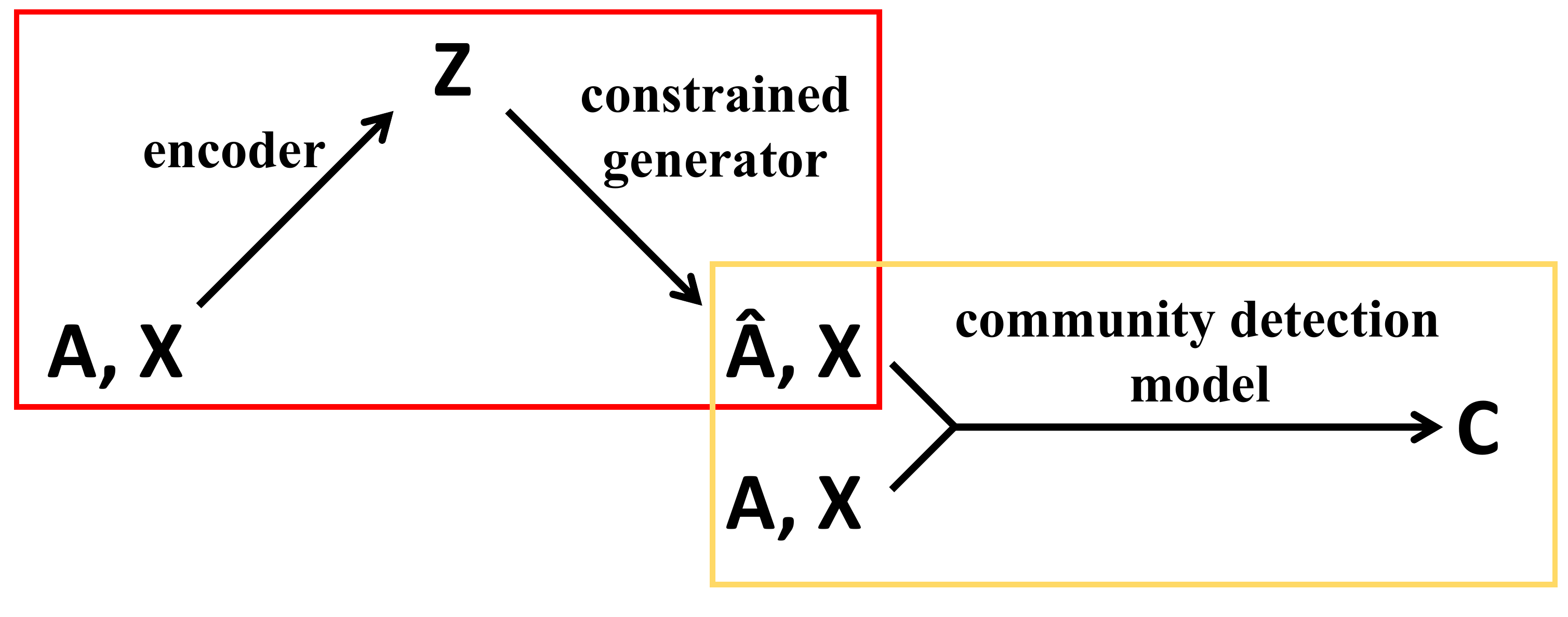}
\end{center}

\caption{Overview of our framework. The adversarial graph generator (in red) produces a constrained adversarial graph, which is used to train a robust surrogate community detection model (in yellow).}
\label{fig.prop}
\end{figure}

In our solution, we design an iterative framework which consists of two neural networks, one working as the adversarial graph generator $g(\cdot)$ and the other working as the surrogate community detection method $f(\cdot)$.  Figure \ref{fig.prop} depicts the framework.  The upper part generates the adversarial graph, which is optimized with respect to the optimum of the surrogate detection model in the lower part.  However, when instantiating the two neural networks, we face three challenges as follows.

\noindent\textbf{Surrogate community detection model}.  How to design a surrogate community detection algorithm, such that the derived adversarial graph also applies to other community detection algorithms?

\noindent\textbf{Imperceptible perturbations}.  What criterion shall we use to ensure the modification is so small that it cannot be perceived by the detection module?

\noindent\textbf{Constrained graph generation}.  How to generate adversarial graphs which meet the imperceptible requirement efficiently?

These three issues make our problem very complicated.  In the following, we present our solutions to the three challenges and then recap our framework at the end of this section.

\subsection{Surrogate Community Detection Model}\label{SCDM}
We propose a novel graph learning based community detection model.  There are two key issues in a community detection model: (1) a distance function to measure the result quality, which corresponds to a loss function in neural networks, and (2) a method to detect communities, which corresponds to the neural network architecture.  We discuss these two issues accordingly below.

\subsubsection{\noindent\textbf{The loss function}}\label{dem}

\begin{figure*}
\begin{center}
\includegraphics [width=0.95\textwidth]{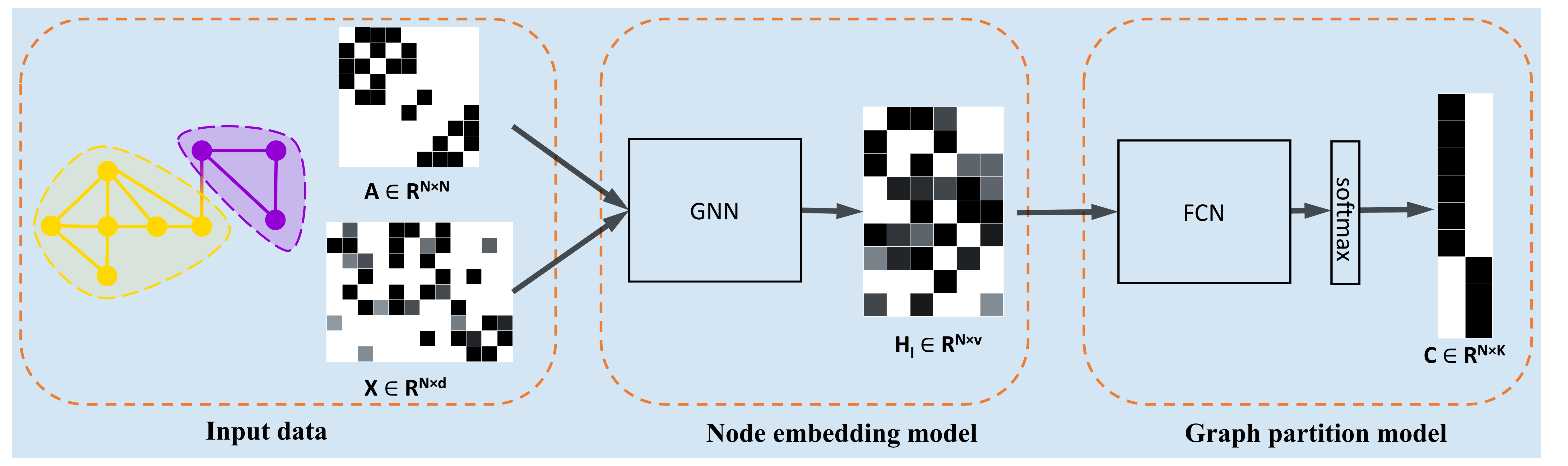}
\label{fig.em}
\end{center}

\caption{Schematic diagram of the surrogate community detection model. It first uses GNNs to derive node embeddings. Based on the node embeddings, it then assigns similar nodes to the same community.}
\label{fig.em}
\end{figure*}

We generalize \emph{normalized cut} \cite{shi2000normalized}, a widely used community detection measure, to serve as the loss function of neural networks.  Normalized cut measures the volume of edges being removed due to graph partitioning:

\begin{equation}
   \frac{1}{K}\sum_{k}\frac{cut(V_k,\overline{V_k})}{vol(V_k)},
\label{equ.H1}
\end{equation}
where $vol(V_k) = \sum_{i \in V_k}degree(i)$, $\overline{V_k} = V\setminus V_k$, $cut(V_k,\overline{V_k}) = \sum_{i \in V_k, j \in \overline{V_k}} A_{ij}$.  The numerator counts the number of edges between a community $V_k$ and the rest of the graph, while the denominator counts the number of incident edges to nodes in $V_k$.  Let $C \in \mathbb{R}^{N\times K}$ be a community assignment matrix where $C_{ik} = 1$ represents node $i$ belongs to community $k$, and $0$ otherwise.  As we focus on detecting non-overlapping communities, an explicit constraint is that $C^\top C$ is a diagonal matrix.  With $C$,  normalized cut can be re-written as:
\begin{equation}
  \frac{1}{K}\sum_{k}\frac{C_{:,k}^\top A(1-C_{:,k})}{C_{:,k}^\top DC_{:,k}},
\label{equ.H2}
\end{equation}
where $D$ is the degree matrix with $D_{ii} = \sum_j A_{ij}$, $C_{:,k}$ is the $k$-th column vector of $C$.  Subtracting Eq.\ \ref{equ.H2} by 1, we get the following simplified target function:
\begin{align}
\begin{split}
  &-\frac{1}{K}\sum_{k}\frac{C_{:,k}^\top AC_{:,k}}{C_{:,k}^\top DC_{:,k}}\\
  =&-\frac{1}{K}\Trace((C^\top AC)\oslash(C^\top DC)),
\end{split}
\end{align}
\label{equ.H3}
where $\oslash$ is element-wise division, $\Trace(\cdot)$ is defined as the sum of elements on the main diagonal of a given square matrix.  Please refer to Appendix \ref{a.c} for detailed derivation for Eqs.\ \ref{equ.H1}-4.

Note that $C^\top C$ needs to be a diagonal matrix, we hereby introduce a new penalization term which explicitly incorporates the constraint on $C$ into the target function.  We subtract $C^\top C$ by $I_K$ as a penalization, where $I_K$ is an identity matrix.  Thus we define the differentiable unsupervised loss function $\mathcal{L}_u$ as follows:
\begin{equation}
  \mathcal{L}_u = -\frac{1}{K}\Trace((C^\top AC)\oslash(C^\top DC)) + \gamma\big|\big|\ \frac{K}{N}C^\top C - I_K\ \big|\big|_F^2,
\label{equ.Hu}
\end{equation}
where $\big|\big|\cdot\big|\big|_F$ represents the Frobenius norm of a matrix, $\frac{K}{N}$ is applied as normalized cut encourages balanced clustering and voids shrinking bias \cite{shi2000normalized,tang2018normalized}. To analyze why the proposed penalization is valid, we suppose there are two communities $i, j$ to be clustered with two assignment vectors $C_{:,i}, C_{:,j}$ respectively.  Suppose each row of $C$ is a distribution and for a non-diagonal element $(C^\top C)_{ij} = \sum_xC_{x, i}C_{x, j}$, it has a maximum value when nodes are assigned uniformly and a minimum value when nodes are assigned discriminatively.  By minimizing this penalization term, we actually encourage nodes to be assigned discriminatively into different communities.

\subsubsection{\noindent\textbf{The network architecture}}\label{ssc}
This part details the neural network architecture to derive the community assignment matrix $C$, which is trained by minimizing the unsupervised loss $\mathcal{L}_u$ in Eq.\ \ref{equ.Hu}. In the literature, spectral clustering \cite{shi2000normalized} is usually used to compute the community assignment matrix $C$.  Due to its limitations in scalability and generalization, some recent work \cite{shaham2018spectralnet} has used a deep learning approach to overcome these shortcomings.  Our architecture, which is based on the recent popular graph neural networks (GNNs) \cite{kipf2017semi,klicpera2018predict} on graph data, has two main parts: (1) node embedding, and (2) community assignment. In the first part, we leverage GNNs to get topology aware node representations, so that similar nodes have similar representations. In the second part, based on the node representations, we assign similar nodes to the same community.  Figure \ref{fig.em} depicts the overall architecture of our community detection model.  In the following, we introduce each part in details.

In this work, we use graph convolutional networks (GCNs) \cite{kipf2017semi} for the purpose of node embedding.  In the preprocessing step, the adjacency matrix $A$ is normalized:
\begin{equation}
  \bar{A} = \tilde{D}^{-\frac{1}{2}}(A + I_N)\tilde{D}^{-\frac{1}{2}},
\label{equ.gene}
\end{equation}
where $I_N$ is the identity matrix and $\tilde{D}_{ii} = \sum_j\ (A + I_N)_{ij}$. Then we transform node features over the graph structure via two-hop smoothing:
\begin{equation}
  H_l = \bar{A} \sigma (\bar{A}XW^0)W^1,
\label{equ.H}
\end{equation}
where $\sigma(\cdot)$ is the activation function such as $ReLU(\cdot)$, $W^0 \in \mathbb{R}^{d \times h}$ and $W^1 \in \mathbb{R}^{h \times v}$ are two weight matrices.  Intuitively, this function can be considered as a Laplacian smoothing operator \cite{jiawww19} for node features over graph structures, which makes nodes more proximal if they are connected within two hops in the graph.  With the node representations $H_l$ in hand, we are now ready to assign similar nodes to the same community by:
\begin{equation}
  C = \textsf{softmax}( \sigma (H_lW_{c1})W_{c2}),
\label{equ.C}
\end{equation}
where $W_{c1} \in \mathbb{R}^{v \times r}$ and $W_{c2} \in \mathbb{R}^{r \times K}$ are two weight matrices.  The function of $W_{c1}$ is to linearly transform the node representations from a $v$-dimensional space to a $r$-dimensional space, then nonlinearity is introduced by tying with the function $\sigma$. $W_{c2}$ is used to assign a score to each of the $K$ communities. Then \textsf{softmax} is applied to derive a standardized distribution for each node over the $K$ communities, which means the summation of the $K$ scores for each node is $1$.

To summarize, we design a neural network to serve as the surrogate community detection model, which is trained in an unsupervised way based on the loss function $\mathcal{L}_u$.  Another function of this module is that it can be used to measure the degree of graph dissimilarities, which is the focus of the next section.

\begin{figure*}
\begin{center}
\includegraphics [width=1\textwidth]{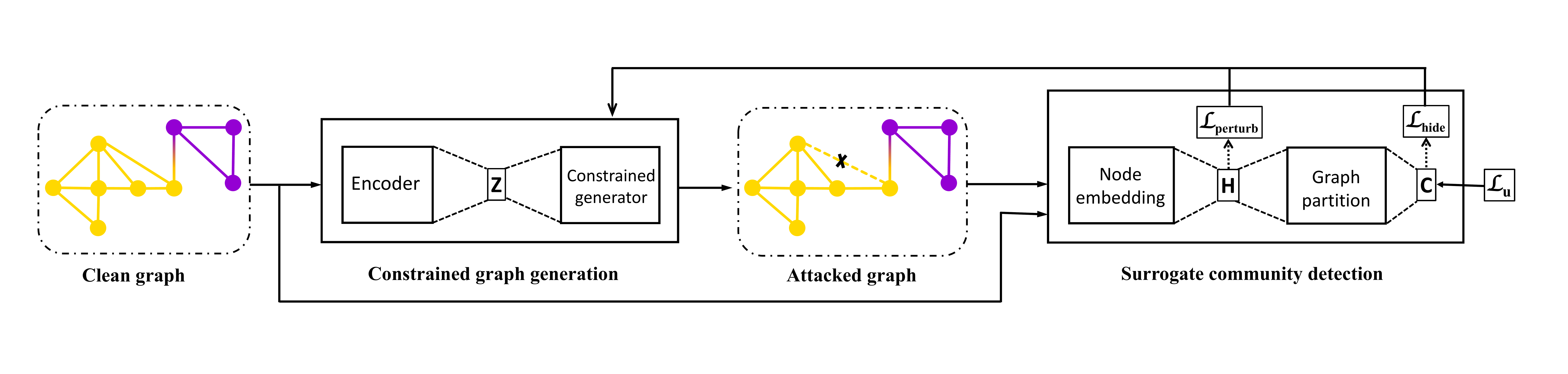}
\end{center}

\caption{The proposed learning framework. There are two modules: constrained graph generation and surrogate community detection. With the optimum of the latter, it will generate two losses to guide the learning of the former.}
\label{fig.sape}
\eat{\vspace{-0.3cm}}
\end{figure*}

\subsection{Imperceptible Perturbations}\label{lrgcn}
An adversarial graph should achieve the goal of potential community detection attack, while, at the same time, it should be as \emph{similar} to the original graph as possible to be stealthy.  In the literature, some work measures the similarity between graphs by the number of discrete changes, for example, \cite{sun2018adversarial, dai2018adversarial} limit the number of allowed changes on edges by a budget $\Delta$:

\begin{equation}
  \sum_{i<j}|A_{ij} - \hat{A}_{ij}| \leq \Delta.
\label{equ.H7}
\end{equation}
Some other work \cite{zugner2018adversarial} argues that the graph dissimilarity should be minimized by maintaining some predefined distribution such as the power-law distribution.  The former based on discrete changes is helpful but far from sufficient, whereas the latter emphasizing a power-law distribution is proven to be rare in reality \cite{broido2019scale}.  In this part, we define a clearly comprehensible graph-related objective to measure the degree of perturbations.

Given $G$ and $\hat{G}$, we know the correspondence between nodes of the two graphs.  The degree of perturbations is measured by the following perturbation loss:
\begin{equation}
\mathcal{L}_{perturb} = \sum_{v_i}KL(ENC(v_i|G) || ENC(v_i|\hat{G})),
\label{equ.ip}
\end{equation}
where $KL(\cdot||\cdot)$ is the Kullback-Leibler divergence, and $KL(P||Q) = \sum_jP_j\log \big(\frac{P_j}{Q_j}\big)$.  Usually $ENC(\cdot)$ learns node representations by capturing some proximities in an unsupervised fashion. We hereby introduce two proximities used to encode node representations:
\begin{itemize}
\item  \emph{Local proximity}: Given node correspondence, the corresponding nodes in two graphs would be similar if their neighbors are similar.
\item  \emph{Global proximity}: Given node correspondence, the corresponding nodes in two graphs would be similar if their relations with all other nodes are similar.
\end{itemize}
The local proximity, or neighborhood/adjacency proximity, is exploited by many node representation methods \cite{perozzi2014deepwalk,kipf2017semi} within a graph.  As we know the node correspondence, it is also valid to be used as a graph similarity measurement \cite{koutra2011algorithms}. By this means, the representations derived in Eq.\ \ref{equ.H} are effective to measure the degree of graph perturbations with respect to the local proximity.

For global proximity, \eat{as it requires to capture the relations with all the other nodes, it immediately suffers from the scalability problem.  In this vein, it is more suitable for small-scale graphs. In this work, we choose to instantiate our global proximity with}we adopt the widely used Personalized PageRank \cite{ilprints422} defined below.
\begin{definition}
Given a starting node distribution $s$, a damping factor $\alpha$, and the normalized adjacency matrix $\bar{A}$, the Personalized PageRank vector $\pi_s$ is defined as:
\begin{equation}
  \pi_s = \alpha s + (1 - \alpha)\pi_s\bar{A}.
\label{equ.ga}
\end{equation}
\end{definition}
With the stationary distribution $\pi_s$, we actually get the influence score of the starting node to all the other nodes.  By taking each node as a starting node, we can get $\mathcal{A}$ in which $\mathcal{A}_{ij}$ denotes the influence score of node $i$ to node $j$.  We can utilize the influence scores by replacing the normalized adjacency matrix $\bar{A}$ in GCN, which leads to the PPNP \cite{klicpera2018predict} formulation:
\begin{equation}
  H_g = \textsf{softmax} (\mathcal{A}XW_g).
\label{equ.H_g}
\end{equation}
There is no overhead for this global proximity based graph perturbation measurement as we can use Eq.\ \ref{equ.H_g} instead of Eq.\ \ref{equ.H} in the surrogate community detection module.  By using global proximity, the graph partition model would be equipped with broader views compared with local proximity used in GCN, which also benefits the community detection module.

\eat{To summarize, in this part we introduce a measurement to quantify the degree of graph perturbations, which is based on two proximities: local and global proximities. }

\subsection{Constrained Graph Generation}\label{CGG}
In this subsection, we describe our method which produces a modified graph $\hat{A}$ that meets the budget constraint in Eq.\ \ref{equ.H7} and the perturbation constraint in Eq. \ref{equ.ip}.  Inspired by the recent success of latent variable generative models such as variational autoencoding \cite{kingma2013auto} in graph generation \cite{kipf2016variational,grover2018graphite}, we use a latent variable model parameterized by neural networks to generate the graph $\hat{A}$.  To be more specific, we focus on learning a parameterized distribution over the adjacency matrix $A$ and feature matrix $X$ as follows:
\begin{equation}
  P(\hat{A}|A,X) = \int q_{\phi}(Z|A,X)p_{\theta}(\hat{A}|A,X,Z)dZ,
\label{equ.ge}
\end{equation}
where $q_{\phi}(Z|A,X)$ is the encoding part and $\phi$ is the encoding parameter set, $p_{\theta}(\hat{A}|A,X,Z)$ is the generation part and $\theta$ is the generation parameter set.  While the encoding part is straightforward and can make use of the corresponding encoders in existing work \cite{kipf2016variational,grover2018graphite}, the generation part is challenging due to the following issues:
\begin{itemize}
\item  \emph{Budget constraint}: Generating a valid graph structure with a budget constraint is hard, as it is a mixed combinatorial and continuous optimization problem.
\item  \emph{Scalability}: Existing graph generation methods suffer from the large-scale problem \cite{wu2019comprehensive}.
\end{itemize}

To satisfy the \emph{budget constraint}, we propose to directly incorporate prior knowledge about the graph structure using the \emph{mask} \cite{kusner2017grammar} mechanism, which can prevent generating certain undesirable edges during the decoding process.  To address the \emph{scalability} issue, we use different solutions based on the input graph size: (1) for a large-scale graph, we design an efficient decoder with $O(m)$ time complexity by focusing on deleting edges, where $m$ is the number of edges, and (2) for a small-scale graph, we are allowed to both delete and insert edges.

\subsubsection{\noindent\textbf{Encoding using GCN}}
We follow VGAE \cite{kipf2016variational} by using the mean field approximation to define the variational family:
\begin{equation}
  q_{\phi}(Z|A,X) = \prod_{i=1}^Nq_{\phi_i}(z_i|A,X),
\label{equ.enc}
\end{equation}
where $q_{\phi_i}(z_i|A,X)$ is the predefined prior distribution, namely, isotropic Gaussian with diagonal covariance.
The parameters for the variational marginals $q_{\phi_i}(z_i|A,X)$ are specified by a two-layer GCN:
\begin{equation}
 \mu,\sigma = GCN_{\phi}(A,X),
\label{equ.H9}
\end{equation}
where $\mu$ and $\sigma$ are the vector of means and standard deviations for the variational marginals $\{q_{\phi_i}(z_i|A,X)\}_{i=1}^N$. $\phi =\{\phi_i\}_{i=1}^N$ is the parameter set for encoding.

\subsubsection{\noindent\textbf{Constrained generation}}
\eat{
Motivated by \cite{kusner2017grammar}, our generator is defined as follows:
\begin{equation}
  p_{\theta}(\hat{A}|A,X,Z) = p_{\theta}(l|A,X,Z)p_{\theta}(\hat{A}|A,X,Z,l)
\end{equation}
where $p_{\theta}(l|A,X,Z)$ is a Poisson distribution.
}
For a large-scale graph $A$, to strictly meet the budget requirement in Eq.\ \ref{equ.H7}, we approximate $p_{\theta}(\hat{A}|A,X,Z)$ by the following operation:
\begin{equation}
p_{\theta}(\hat{A}|A,X,Z)= \prod_{(i,j) \in S}\Theta(E_{ij}),
\end{equation}
where $E_{ij} = [Z_i|X_i]\odot[Z_j|X_j]$, $\odot$ represents element-wise multiplication, $Z_i|X_i$ is a concatenation of $Z_i$ and $X_i$, $\Theta(\cdot)$ and $S$ are defined as below:
\begin{equation}
  b_{ij} = W_{b1}\sigma(W_{b2}E_{ij})\ \ if\ \ \ A_{ij}=1,
\label{H5}
\end{equation}
\eat{
\begin{equation}
  S = \{(i,j):b_{ij} \in \textsf{top} (\{b_{ij}\}_1^m,m-\Delta)\},
\label{equ.HT}
\end{equation}

\begin{equation}
  b_{ij} = 0\ \ \ \ \ if\ \ \ \ (i,j) \notin S,
\label{Ht}
\end{equation}
}
\begin{equation}
  \Theta(E_{ij})= \frac{e^{b_{ij}}}{\sum e^{b_{ij}}}\ \ \ \ \ \ if\ \ \ \ A_{ij}=1,
\label{H6}
\end{equation}
where $\Theta(E_{ij})$ is the computed score for keeping the edge by two-layer perceptrons and \textsf{softmax} function.  We sample without replacement $m-\Delta$ edges according to their keeping scores $\Theta(\cdot)$ defined in Eq. \ref{H6}, and we use set $S$ to denote these edges that have been chosen. Intuitively, we want to select $m-\Delta$ edges that exist in the original graph and maximize their product of keeping scores. With this strategy, we can also generate graphs that strictly meet the budget requirement in Eq.\ \ref{equ.H7}.

\eat{By $\textsf{top} (\{b_{ij}\}_1^m,m-\Delta)$, we sort them in descending order and add the top $m - \Delta$ edges into set $S$.} For a small-scale graph, we split the budget $\Delta$ into two parts: $\Delta/2$ for deleting edges and $\Delta/2$ for inserting edges. We approximate $p_{\theta}(\hat{A}|A,X,Z)$ by:
\begin{align}
\begin{split}
p_{\theta}(\hat{A}|A,X,Z) = \prod_{(i,j) \in S}\Theta(E_{ij})\prod_{(i,j) \in \bar{S}}\Psi(E_{ij}),
\end{split}
\end{align}
where $S$ and $\Theta(\cdot)$ are derived in the same way as Eqs.\ \ref{H5}-\ref{H6} except that we just mask out or delete $\Delta/2$ edges.  $\bar{S}$ denotes the set of edges that have been chosen to be inserted.  $\Psi(\cdot)$ is computed in the same way as $\Theta(\cdot)$ with non-identical parameters and the essential differences are: (1) we compute with the condition of $A_{ij} = 0$, and (2) we add $\Delta/2$ edges to $\bar{S}$.

\subsubsection{\noindent\textbf{The loss function}}
Different from the traditional VAE, our method is not optimized by the unsupervised reconstruction error, rather it is powered by the following combined loss:
\begin{align}
\begin{split}
  \mathcal{L}_g = \mathcal{L}_{prior} + &(\lambda_1\mathcal{L}_{hide} + \lambda_2\mathcal{L}_{perturb})(\sum_{(i,j) \in S}\log \Theta(E_{ij}) \\
  + &\sum_{(i,j) \in \bar{S}}\log \Psi(E_{ij})),
\label{equ.Hz}
\end{split}
\end{align}
where $\mathcal{L}_{prior} = KL(q(Z|X,A)||p(Z))$ is the prior loss with $p(Z) = \prod_{i=1}^NP(z_i)=\prod_{i=1}^N\mathcal{N}(z_i|0,I)$, $\mathcal{L}_{perturb}$ is the imperceptible perturbation requirement introduced in Eq.\ \ref{equ.ip}, $\mathcal{L}_{hide}$ is used to diverge the community assignment distributions within the set of individuals:
\begin{equation}
  \mathcal{L}_{hide} = \min_{i \in C^+,j \in C^+} KL(C_{i,:}||C_{j,:}),
\label{equ.H8}
\end{equation}
where $\mathcal{L}_{hide}$ can be regarded as the margin loss or the smallest distance for any pair inside $C^+$, $\lambda_1 < 0$ as we aim to maximize the margin so that the members of $C^+$ are spread out across the communities in $\cup_{i=1}^KG_i$.

To summarize, Eq. \ref{equ.Hz} receives error signals $\mathcal{L}_{hide}$ and $\mathcal{L}_{perturb}$ from the surrogate community detection model, which serves as a \noindent\textbf{reward} to guide the optimization of our graph generator.
\subsection{The Proposed Model}\label{CGG}
From the perspective of Actor-Critic framework, the adversarial attacker $g(\cdot)$ could be considered as the Actor and the community detection algorithm $f(\cdot)$ could be regarded as the Critic. As the error signals of adversarial graph generator are obtained from the community detection model and the community detection model needs the generated graphs as inputs for robust training, we design an iterative framework, named \underline{C}ommunity \underline{D}etection \underline{ATTACK}er (CD-ATTACK), to alternate between minimizing the loss of both $g(\cdot)$ and $f(\cdot)$. We refer to Figure \ref{fig.sape} and Algorithm \ref{algo.set} for details of the training procedure.

At the beginning of Algorithm \ref{algo.set}, we exploit the constrained graph generator $g(G, \Delta)$ so as to get an adversarial graph $\hat{G}$ (line 4).  We then utilize the idea of robust training and feed both $G$ and $\hat{G}$ into the surrogate community detection model $f(\cdot)$ to compute $\mathcal{L}_{u}$ (line 6). Based on the optimum of $f(\cdot)$, we get $\mathcal{L}_{perturb}$ and $\mathcal{L}_{hide}$ to power the learning process of $g(\cdot)$ (line 7-8).

In practice, we have observed the devil in the training process.  We therefore provide a list of practical considerations:
\begin{itemize}
\item  \emph{Pre-training community detection model}: An appealing property of CD-ATTACK is that pre-training the community detection model before the iteration results in better performance and faster convergence.
\item  \emph{Normalization trick of $G$}: When feeding both $G$ and $\hat{G}$ into $f(\cdot)$, there could be trouble as $G$ and $\hat{G}$ have discrete differences. We thus observe better results if decoupling self-loop and neighborhood smoothing in GCN, i.e., $\bar{A} = \tilde{D}^{-\frac{1}{2}}A\tilde{D}^{-\frac{1}{2}}$ and another weight matrix for self-loop.
\end{itemize}

\begin{algorithm}
  \caption{Training the model}
  \label{algo.set}
  \KwIn{$G$, $\Delta$.}
  \KwOut{$\hat{G}$.}
  Initial: parameters $\mathcal{W}_{g}$,$\mathcal{W}_{f}$;

  \Repeat{deadline}{
    $G \leftarrow$ full batch\;
	$\hat{G} \leftarrow$ $g(G,\Delta)$\;
    $\mathcal{L}_{prior} \leftarrow KL(q(Z|X,A)||p(Z))$\;
	$\mathcal{L}_{u} \leftarrow f(\{G,\hat{G}\})$\;
	$\mathcal{L}_{perturb} \leftarrow \sum_{v_i}KL(ENC(v_i|G) || ENC(v_i|\hat{G}))$\;
	$\mathcal{L}_{hide} \leftarrow \min_{i \in C^+,j \in C^+} KL(C_{i,:}||C_{j,:})$\;
    // Update parameters according to gradients

    $\mathcal{W}_{g} \leftarrow^+ -\bigtriangledown_{\mathcal{W}_{g}} \mathcal{L}_g$\;
    $\mathcal{W}_{f} \leftarrow^+ -\bigtriangledown_{\mathcal{W}_{f}} \mathcal{L}_{u}$\;
  }
\end{algorithm}

\begin{table}
  \caption{Statistics of graphs considered}
  \label{tab:twodatasets}
  \scalebox{0.9}{
  \begin{tabular}{ccccc}
    \toprule
& No. of nodes&No. of edges&No. of features \\
    \midrule
	\textbf{DBLP-medium}&5,304&28,464&305\\
	\textbf{Finance-medium} &5,206&5,494&7\\
	\textbf{DBLP-large}&20,814&119,854&305\\
	\textbf{Finance-large} &20,121&23,732&7\\
  \bottomrule
\end{tabular}
}
\end{table}

\begin{table*}
  \caption{Performance comparison of different attacks on the surrogate model}
  \begin{tabular}{ccccccccc}
    \toprule
Data sets&\multicolumn{2}{c}{\textbf{DBLP-medium}}&\multicolumn{2}{c}{\textbf{Finance-medium}}&\multicolumn{2}{c}{\textbf{DBLP-large}}&\multicolumn{2}{c}{\textbf{Finance-large}}\\
\hline
       -      &M1&M2&M1&M2&M1&M2&M1&M2\\
    \midrule
	\textbf{DICE} &6.35\%&41.44\%&3.25\%&53.55\%&6.91\%&50.77\%&3.66\%&72.11\%\\
	\textbf{MBA} &7.33\%&44.88\%&2.80.\%&52.00\%&7.04\%&53.67\%&3.08\%&72.11\%\\
	\textbf{RTA} &6.55\%&42.22\%&2.53\%&42.66\%&6.25\%&44.66\%&1.74\%&55.11\%\\
	\hline
	\textbf{CD-ATTACK} &\textbf{13.72\%}&\textbf{52.00\%}&\textbf{3.32\%}&\textbf{63.00\%}&\textbf{8.11\%}&\textbf{57.62\%}&\textbf{4.14\%}&\textbf{87.33\%}\\
  \bottomrule
\end{tabular}
 \label{alll}
\vspace{-0.4cm}
\end{table*}

\section{EXPERIMENTS}\label{sec.exp}
We validate the effectiveness of our model on four real-world data sets: (1) DBLP-medium, (2) Finance-medium, (3) DBLP-large, and (4) Finance-large.  We first evaluate the surrogate model, then check if the attack can be transferred to other community detection models.

\subsection{Data}
\subsubsection{DBLP}\footnote{\url{https://dblp.uni-trier.de/}}
From DBLP bibliography data, we build two coauthor graphs with 5,304 and 20,814 authors respectively.  The former is named \emph{DBLP-medium} and the latter is named \emph{DBLP-large}.  For each author, we use a one-hot representation with 305 dimensions encoding his/her research keywords. Accordingly we construct an adjacency matrix $A$ by denoting $A_{ij} = 1$ and $A_{ji} = 1$ if the two authors have coauthored papers.

\subsubsection{Finance}
From an anonymized user-user transaction data set provided by Tencent, we select 5,206 and 20,121 users and build two transaction networks respectively.  The former is named \emph{Finance-medium} and the latter is named \emph{Finance-large}. For each user, we collect 7 features.  We construct an adjacency matrix $A$ by denoting $A_{ij} = 1$ and $A_{ji} = 1$ if the two users have one or more transaction records.

Table \ref{tab:twodatasets} lists the statistics of the four graphs.

\subsection{Baselines and Metrics}\label{syn.base}
\subsubsection{Baselines}
We use the following approaches as our baselines:
\begin{itemize}
\item DICE \cite{waniek2018hiding}, which is a heuristic attack strategy. It first deletes some edges incident to the target set $C^+$, then spends the remaining budget for inserting edges between $C^+$ and the rest of the graph.

\item Modularity Based Attack (MBA) \cite{fionda2017community}, which weakens the community structure by deleting intra-community edges and inserting inter-community edges. It is based on modularity \cite{newman2006modularity}.

\item Random Target Attack (RTA), which follows the idea of RND \cite{zugner2018adversarial}. Given $C^+$, in each step we randomly sample a node.  If the node is connected to $C^+$, we delete an edge between the node and $C^+$ randomly; otherwise we add an edge between the node and $C^+$ randomly.

\end{itemize}

\subsubsection{Metrics}

We follow \cite{waniek2018hiding} and use two measures to quantify the degree of hiding:

\begin{equation}
M1(C^+,G) = \frac{|{G_i:G_i\cap C^+ \neq \emptyset}| - 1}{(K -1)\times \max_{G_i}(|G_i\cap C^+|)},
\end{equation}
where $K > 1$. The numerator grows linearly with the number of communities that $C^+$ is distributed over.  The denominator penalizes the cases in which $C^+$ is skewly distributed over the community structures. Intuitively this measure $M1(C^+,G)$ focuses on how well the members of $C^+$ spread out across $G_1, \ldots, G_K$ and $M1(C^+,G) \in [0,1]$.
\begin{equation}
M2(C^+,G) = \sum_{G_i:G_i\cap C^+ \neq \emptyset}\frac{|G_i \setminus C^+|}{\max(N-|C^+|,1)},
\end{equation}
where the numerator grows linearly with the number of non-members of $C^+$ that appear simultaneously with members of $C^+$ in the same community. Note the numerator only counts the communities in which there exists a member of $C^+$. Basically, this measure focuses on how well $C^+$ is hidden in the crowd and $M2(C^+,G) \in [0,1]$.

For both measures, a greater value denotes a better hiding performance.
\subsubsection{Setup}
For all the data sets, we set $K = 10$, i.e., we cluster the nodes in each graph into 10 communities. We select the target nodes by the following process: (1) use Infomap \cite{rosvall2008maps} to divide the graph into 10 communities, and (2) in each community select 5 nodes with the highest degrees and another $5$ nodes randomly.  Thus we totally have $(5+5)\times 10 = 100$ target nodes for our attack. The budget $\Delta$ is always 10, if not specified otherwise.

For CD-ATTACK, We use the same network architecture through all the experiments. Our implementation is based on Tensorflow.  We train the model using minibatch based Adam optimizer with exponential decay. We set $\gamma = 0.1$, the output dimension of the first-layer GCN to be 32 and that of the second-layer GCN to be 16.  In addition, we set the fully connected layer with 32 rectified linear units with a dropout rate of 0.3. The initial learning rate is 0.001.

For fair comparison, we separate the training process of CD-ATTACK and the surrogate community detection model. In other words, CD-ATTACK receives no feedback from the model to be attacked. We run all the methods 5 times and report their average performance.
\subsection{Attacks on the Surrogate Model}

\subsubsection{Attack performance}\label{xix}
Table \ref{alll} lists the experimental results on the four data sets.  As all the attacks are based on the same graphs in each data set, we report the values of $M1$ and $M2$ rather than the gain/changes of these measures.  Among all approaches, CD-ATTACK achieves the best performance on all data sets in terms of both measures, which demonstrates CD-ATTACK's superiority.  In the following, we compare the performance of the three baseline methods.

\noindent\underline{DICE}:  DICE performs better than RTA on all data sets except \emph{DBLP-medium}, worse than MBA on most data sets except \emph{Finance-medium}.  One possible explanation is that DICE follows the heuristic that decreasing the density of the connections with $C^+$ will help, while RTA is randomly based and it has a higher probability not to cut the edges between targets.

\noindent\underline{MBA}: MBA performs quite well compared with the other two baselines, which proves the effectiveness of adopting \emph{modularity} to select the candidate edges. However it performs worse than DICE on \emph{Finance-medium}, meaning it is not stable.

\noindent\underline{RTA}: RTA performs worse than most methods except that it outperforms DICE on \emph{DBLP-medium}. A possible explanation is that randomness may win in some scenarios as it can explore more possibilities.

\subsubsection{Effect of the budget $\Delta$}
\begin{figure}
\begin{center}
\includegraphics [width=0.35\textwidth,scale=1]{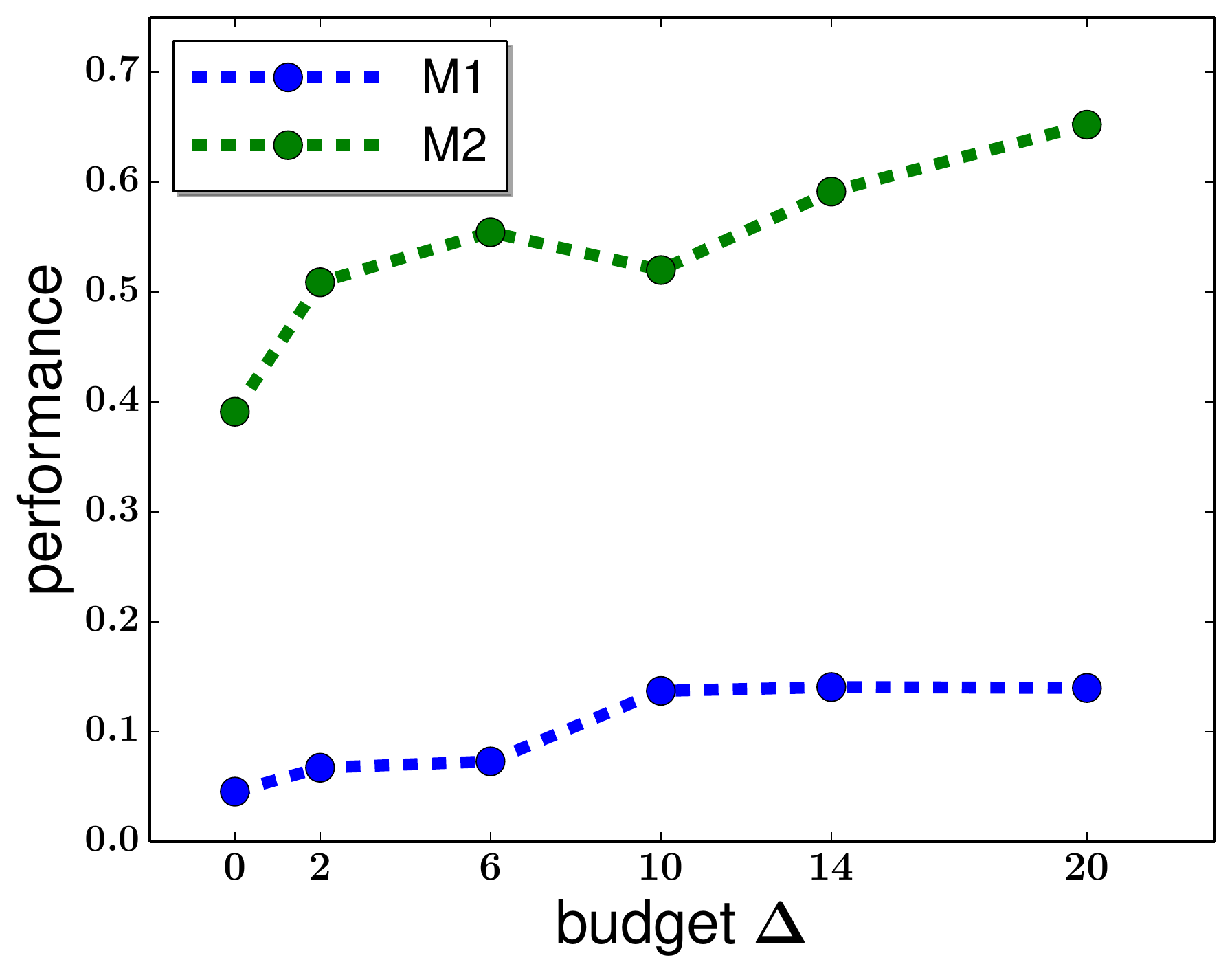}
\end{center}

\caption{CD-ATTACK performance with different budgets on \emph{DBLP-medium}}
\label{fig.delta}
\end{figure}

We evaluate how the budget $\Delta$ affects the attack performance.  Taking CD-ATTACK on \emph{DBLP-medium} as an example, we vary $\Delta$ from 2 to 20 and plot the corresponding attack performance in terms of $M1$ and $M2$ in Figure \ref{fig.delta}.  \eat{Note that $M1$ and $M2$ reported here are the orignal values rather than the gain before-and-after the attack.} As we increase $\Delta$, the attack performance improves (i.e., both $M1$ and $M2$ increase), meaning that a larger budget is helpful for a successful attack.  Another observation is that the attack performance increases quite fast when $\Delta\leq 10$ and becomes stable when $\Delta>10$, indicating the effect of $\Delta$ becomes less obvious as we further increase the budget.

\subsubsection{Adversarial cost}
\begin{figure}
\begin{center}
\includegraphics [width=0.48\textwidth,scale=1]{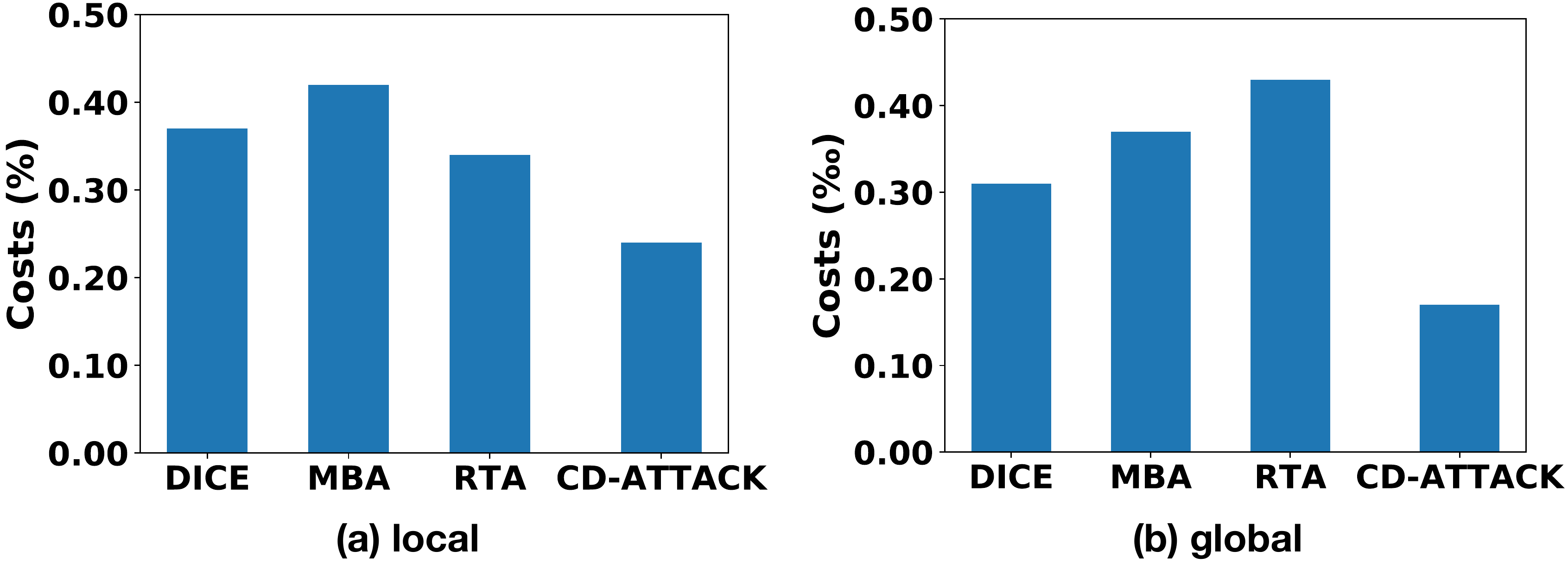}
\end{center}

\caption{Cost comparison of different attacks on \emph{DBLP-medium}}
\label{fig.cost}
\end{figure}

We compare the adversarial costs of different attacks on \emph{DBLP-medium}.  The result is shown in Figure \ref{fig.cost}, in which \emph{local} means we use GCN for the neighborhood smoothing in the surrogate community detection model and \emph{global} means we use PPNP \cite{klicpera2018predict} instead.  From the figure we can find that our method achieves the lowest perturbation loss, and this is not surprising as we explicitly take this constraint into consideration when selecting the proper candidates.

\subsubsection{Visualization}

\begin{figure*}
\begin{center}
\includegraphics [width=0.8\textwidth,scale=1]{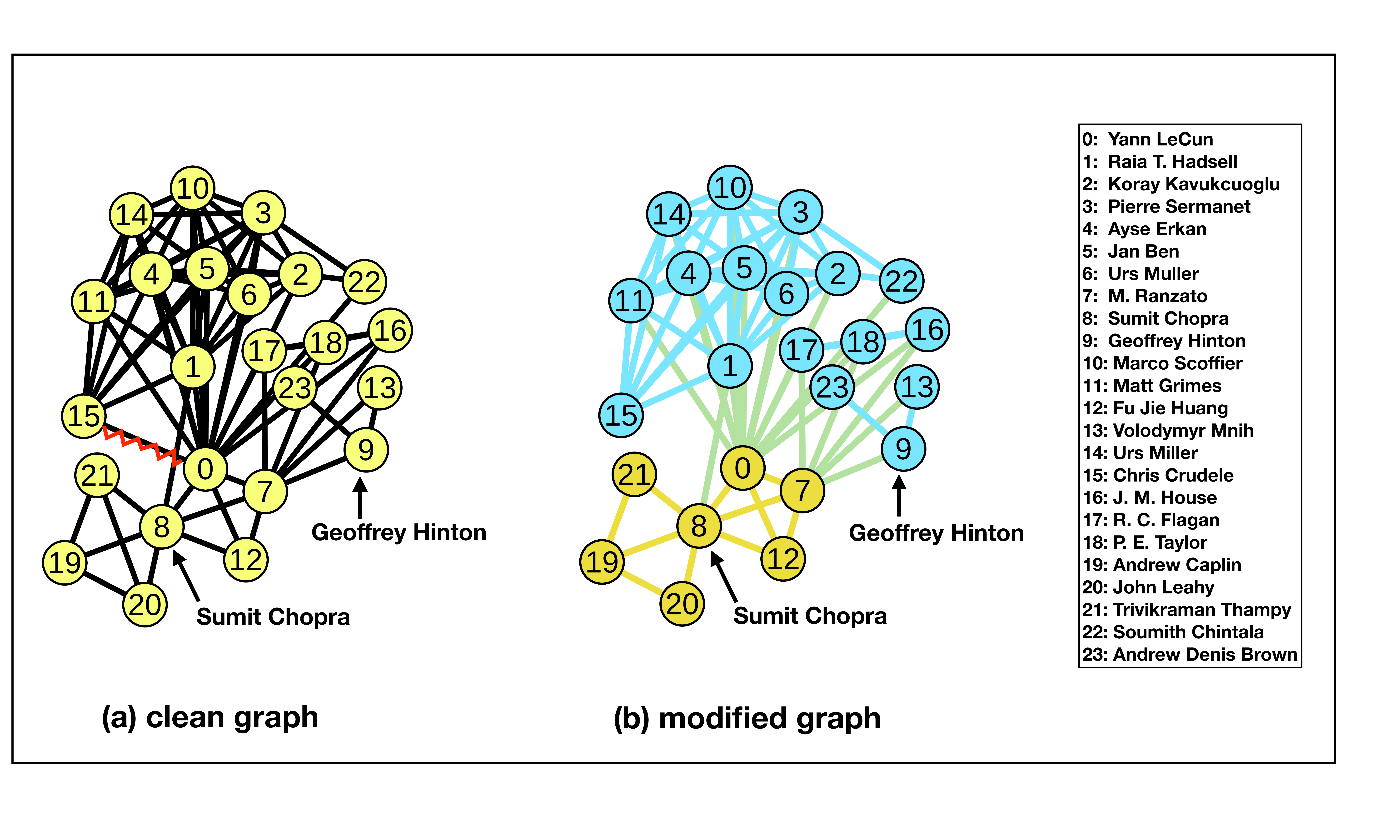}
\end{center}

\caption{A case to show how CD-ATTACK hides individuals}
\label{fig.va}
\end{figure*}

To have a better understanding of how CD-ATTACK works, we target a subgraph of \emph{DBLP-medium} which consists of 24 nodes under the attack in Section \ref{xix}.  We have two target nodes 8 and 9 included in this subgraph, which correspond to \emph{Sumit Chopra} and \emph{Geoffrey Hinton} respectively. We inspect how the community assignment changes with respect to the attack, as depicted in Figure \ref{fig.va}.  In this figure, different colors represent different community assignment.  As we can see, at the beginning the community detection algorithm considers the two targets are in the same community (on the left). As we cut an edge $(0, 15)$ (in red), the community detection algorithm separates the subgraph into two parts, thus fails to treat the two targets in one community.  An interesting observation is that the attack CD-ATTACK chooses is not a direct attack but rather an influence attack, as the removed edge does not involve either node 8 or 9 as an end point, which differs from most of our competitors.

\eat{
\begin{table}
  \caption{Cost comparison of different attacks on DBLP-medium}
  \label{tta}
  \begin{tabular}{ccc}
    \toprule
  \textbf{Algorithm}&\textbf{local}&\textbf{global} \\
    \midrule
   \textbf{DICE} & 0.37\% & 0.31\%  \\
   \textbf{MBA} &0.42 \% & 0.37 \% \\
   \textbf{RTA} & 0.34 \% & 0.43 \% \\
   \hline
   \textbf{CD-ATTACK (local)} & 0.24 \% & -\\
   \textbf{CD-ATTACK (global)} & - & \textbf{0.17}  \% \\
	  \bottomrule
\end{tabular}
\eat{\vspace{-0.4cm}}
\end{table}
}

\subsection{Transferability}
In this part, we shall explore whether the adversarial graphs generated by our methods also apply to other graph learning based community detection models. In this vein, we select two widely used target models as follows:
\begin{itemize}
\item Node2vec + K-means (NK) which first uses Node2vec \cite{grover2016node2vec} to get node embeddings and then utilizes K-means to derive community assignments.
\item ComE \cite{cavallari2017learning} which jointly solves community embedding, community detection and node embedding in a closed loop. We use the highest probability across the $K$ communities as the last community assignment.
\end{itemize}
We test the transferability of our method with the following procedure: (1) run our method and three competitors and get the corresponding adversarial graphs, and (2) train the target models based on the adversarial graphs. We report the performance on \emph{DBLP-medium} and \emph{Finance-medium} in Table \ref{tel} and Table \ref{telem} for NK and ComE respectively.

As we can see, while all the attacking methods are effective, our method achieves the best transferability on most measures. We therefore conclude that our surrogate model and  \emph{normalized cut} measure are a sufficient approximation of the true measures used in other models.

\begin{table}
  \caption{Performance comparison of different attacks on NK}
  \label{tel}
  \begin{tabular}{ccccc}
    \toprule
  Data sets&\multicolumn{2}{c}{\textbf{DBLP-medium}}&\multicolumn{2}{c}{\textbf{Finance-medium}}\\
\hline
       -      &M1&M2&M1&M2\\
    \midrule
   \textbf{DICE} & 3.05\% & 21.11\% & 12.78\%&  23.70\%\\
   \textbf{MBA} &2.12 \% & 8.77 \% &\textbf{14.44}\%&18.15\%\\
   \textbf{RTA} & 2.54 \% & 8.33 \% &12.78\%&18.15\%\\
   \hline
   \textbf{CD-ATTACK} & \textbf{4.04} \% & \textbf{21.77} \% &13.89\%&\textbf{33.34}\%\\
	  \bottomrule
\end{tabular}
\eat{\vspace{-0.4cm}}
\end{table}

\begin{table}
  \caption{Performance comparison of different attacks on ComE}
  \label{telem}
  \begin{tabular}{ccccc}
    \toprule
  Data sets&\multicolumn{2}{c}{\textbf{DBLP-medium}}&\multicolumn{2}{c}{\textbf{Finance-medium}}\\
\hline
       -      &M1&M2&M1&M2\\
    \midrule
   \textbf{DICE} & 3.08\% & 5.33\% & 5.56\%& 45.00\%\\
   \textbf{MBA} &1.58 \% & 5.18 \% & \textbf{10.00}\%& 41.67\%\\
   \textbf{RTA} & 3.14 \% & 4.77 \% & 4.44\%& 28.33\%\\
   \hline
   \textbf{CD-ATTACK} & \textbf{5.56}\% & \textbf{7.33}\% & 5.56\%& \textbf{65.00}\%\\
	  \bottomrule
\end{tabular}
\eat{\vspace{-0.4cm}}
\end{table}

\section{Related Work}\label{sec.related}
\eat{This work is related to prediction tasks in time-evolving graphs, path representation and neural network on graphs.}

\noindent\textbf{Attacks on Community Detection}.  In the literature, three studies have a similar setting as ours in the sense that they also want to hide a set of individuals. \cite{nagaraja2010impact} first formulates the problem and comes up with a solution by adding edges for nodes with high degree centrality.  Later \cite{waniek2018hiding} proposes two heuristic algorithms ROAM and DICE for hiding an individual and a community respectively. It also formulates two measures to quantify the level of deception of a detection algorithm, which we follow in this work. \cite{fionda2017community} further extends the idea by proposing the concept of community deception and two attacking methods: safeness-based deception and modularity-based deception. Our work differs from them in three aspects: (1) we focus on deep graph based community detection methods, (2) our method can handle attributed graphs while none of the previous studies can, and (3) we consider not only a budget but also a graph similarity to ensure unnoticeable perturbations.

\noindent\textbf{Neural Network based Community Detection}.  There have been many neural network based community detection methods ever since the pioneer graph embedding studies \cite{perozzi2014deepwalk,grover2016node2vec}.  One straightforward approach is first utilizing node embedding methods such as Node2vec \cite{grover2016node2vec} and then applying K-means to obtain cluster assignment results.  However, such a pipeline lacks a unified optimization objective, which makes the results suboptimal.  In this regard, \cite{cavallari2017learning} jointly solves community embedding, community detection and node embedding in a closed loop.  However, the gradient still cannot be passed between different modules. \cite{chen2017supervised} overcomes this problem by introducing a line-graph based variation of GNNs in a supervised community detection scenario. \cite{nazi2019gap} further generalizes GNNs to tackle unsupervised community detection problems based on min-cut. Our design is similar to \cite{nazi2019gap} except that we encourage cluster assignments to be orthogonal by a novel regularization.

\noindent\textbf{Imperceptible Perturbations}.  Adversarial perturbations are required to be imperceptible in order to foul the corresponding detecting module.  In the image domain, one can use $L_p$ norm distance \cite{carlini2017towards} to achieve unnoticeable modifications.  In the graph domain, it is still an open problem to ensure unnoticeable perturbations \cite{sun2018adversarial}.  In \cite{dai2018adversarial}, the attacker is restricted to add/remove edges in the original graph within a budget. In \cite{zugner2018adversarial}, the attacker is further required to maintain the power-law degree distribution in addition to the budget constraint.  However, maintaining the power-law distribution is neither necessary nor sufficient as a recent study \cite{broido2019scale} has found real-world networks with a power-law distribution are very rare.  In this work, we leverage the general graph similarity metrics to measure the degree of modifications, in the hope of guaranteeing minimal perturbations on the original graph. Especially, we instantiate our measurements by two similarities: Personalized PageRank \cite{ilprints422} and adjacency matrix based similarity.

\noindent\textbf{Constrained Graph Generation}.  Although there are many studies on graph generation~\cite{kipf2016variational,grover2018graphite}, constrained graph generation has been studied less.  GVAE \cite{kusner2017grammar} first constructs parse trees based on the input graph and then uses Recurrent Neural Networks (RNNs) to encode to and decode from these parse trees.  It utilizes the binary \emph{mask} mechanism to delete invalid vectors. NeVAE \cite{samanta2019nevae} and CGVAE \cite{liu2018constrained} both leverage GNNs to generate graphs which match the statistics of the original data. They make use of a similar \emph{mask} mechanism to forbid edges that violate syntactical constraint. Our model also uses the \emph{mask} mechanism to prevent generating undesirable edges. \eat{One difference between our generation model and others is that we replace the reconstruction loss by a newly proposed graph similarity loss, which is more suitable as the generated graph and the original graph have a discrete difference in the number of edges.}

\noindent\textbf{Relation with Other Frameworks}.  We design an iterative framework in this work. It has some connections with some popular frameworks.
\begin{itemize}
\item Generative Adversarial Network (GAN) \cite{goodfellow2014generative}, which is an iterative framework consisting of two modules: the generator and the discriminator. Usually GAN is formulated as a \emph{minimize/maximize} paradigm in which the generator aims to minimize the distance between true and generated data, while the discriminator wants to maximize the distance. Our framework is different as it does not belong to this paradigm.
\item Actor-Critic framework \cite{konda2000actor}, which is a class of techniques in reinforcement learning. It consists of two modules: the actor (generator) and the critic (discriminator).  Usually the actor must learn based on the estimated error signals, e.g., policy gradient, from the critic. Our model belongs to this framework.
\end{itemize}

\section{CONCLUSION}\label{sec.con}

In this paper, we study adversarial attack on graph learning based community detection models via hiding a set of nodes.  To overcome the difficulty of no explicit feedback from the detection module in reality, we design a new graph learning based community detection model based on the widely used GNNs and normalized cut.  Given the huge search space of selecting proper candidate edges to change, we take a learning approach and design a graph generator that can meet the discrete constraint.  We systematically analyze how we can measure the adversarial costs of various attacks and come up with a clear graph objective from two perspectives: local proximity and global proximity.  Experimental results on four real-world data sets show that CD-ATTACK outperforms other competitors by a significant margin in two measures.  The adversarial graphs generated can also be transferred to other graph learning based community detection models.

\begin{acks}
The work described in this paper was supported by grants from the Research Grant Council of the Hong Kong Special Administrative Region, China [Project No.: CUHK 14205618], Tencent AI Lab Rhino-Bird Focused Research Program GF201801.
\end{acks}


\appendix
\section{Derivations of equations (2), (3) and (4)}\label{a.c}

1. The derivation from Eq.\ 2 to Eq.\ 3.

Firstly, we have
\begin{equation}
\label{eq:cut}
\begin{split}
cut(V_k, \bar{V}_k) &= \sum_{k'\neq k}C_{:, k}^T A C_{:, k'}\\
&= C_{:, k}^T A (\sum_{k'\neq k} C_{:, k'})\\
&= C_{:, k}^T A (\bm{1} - C_{:, k})
\end{split}
\end{equation}
where $\bm{1}$ is the vector with ones. Besides, we have

\begin{equation}
\label{eq:vol}
vol(V_k) = C_{:, k}^T D C_{:, k}
\end{equation}

Based on Eqs.\ \ref{eq:cut} and \ref{eq:vol}, Eq.\ 2 can be re-written into Eq.\ 3 as follows:

\begin{equation}
\begin{split}
Ncut &= \frac{1}{K} \sum_{k}\frac{cut(V_k, \bar{V}_k)}{vol(V_k)}\\
&= \frac{1}{K} \sum_{k}\frac{C_{:, k}^T A (\bm{1} - C_{:, k})}{C_{:, k}^T D C_{:, k}}
\end{split}
\end{equation}

2. The derivation from Eq.\ 3 to Eq.\ 4.

Firstly, let us look at the main part in Eq.\ 3.
\begin{equation}
\begin{split}
\frac{C_{:, k}^T A (\bm{1} - C_{:, k})}{C_{:, k}^T D C_{:, k}} &= \frac{C_{:, k}^T A \bm{1} - C_{:, k}^T A C_{:, k}}{C_{:, k}^T D C_{:, k}}\\
&= \left( 1 - \frac{C_{:, k}^T A C_{:, k}}{C_{:, k}^T D C_{:, k}}\right )
\end{split}
\end{equation}

So the normalized cut loss can be re-written as
\begin{equation}
\begin{split}
Ncut &= \frac{1}{K} \sum_{k}\frac{C_{:, k}^T A (\bm{1} - C_{:, k})}{C_{:, k}^T D C_{:, k}}\\
&= \frac{1}{K} \sum_{k} \left( 1 - \frac{C_{:, k}^T A C_{:, k}}{C_{:, k}^T D C_{:, k}}\right )\\
&= 1 - \frac{1}{K} \sum_{k}  \frac{C_{:, k}^T A C_{:, k}}{C_{:, k}^T D C_{:, k}}\\
&= 1 - \frac{1}{K} \sum_{k}  \frac{(C^T A C)_{k,k}}{(C^T D C)_{k,k}}\\
&= 1 - \frac{1}{K} Tr\left ( (C^T A C) \oslash (C^T D C) \right )\\
\end{split}
\end{equation}

\bibliographystyle{ACM-Reference-Format}
\bibliography{sample-bibliography}


\begin{thebibliography}{37}


\ifx \showCODEN    \undefined \def \showCODEN     #1{\unskip}     \fi
\ifx \showDOI      \undefined \def \showDOI       #1{#1}\fi
\ifx \showISBNx    \undefined \def \showISBNx     #1{\unskip}     \fi
\ifx \showISBNxiii \undefined \def \showISBNxiii  #1{\unskip}     \fi
\ifx \showISSN     \undefined \def \showISSN      #1{\unskip}     \fi
\ifx \showLCCN     \undefined \def \showLCCN      #1{\unskip}     \fi
\ifx \shownote     \undefined \def \shownote      #1{#1}          \fi
\ifx \showarticletitle \undefined \def \showarticletitle #1{#1}   \fi
\ifx \showURL      \undefined \def \showURL       {\relax}        \fi
\providecommand\bibfield[2]{#2}
\providecommand\bibinfo[2]{#2}
\providecommand\natexlab[1]{#1}
\providecommand\showeprint[2][]{arXiv:#2}

\bibitem[\protect\citeauthoryear{Ahn, Bagrow, and Lehmann}{Ahn
  et~al\mbox{.}}{2010}]%
        {ahn2010link}
\bibfield{author}{\bibinfo{person}{Yong-Yeol Ahn}, \bibinfo{person}{James~P
  Bagrow}, {and} \bibinfo{person}{Sune Lehmann}.}
  \bibinfo{year}{2010}\natexlab{}.
\newblock \showarticletitle{Link communities reveal multiscale complexity in
  networks}.
\newblock \bibinfo{journal}{\emph{Nature}} \bibinfo{volume}{466},
  \bibinfo{number}{7307} (\bibinfo{year}{2010}), \bibinfo{pages}{761--764}.
\newblock


\bibitem[\protect\citeauthoryear{Akoglu, Tong, and Koutra}{Akoglu
  et~al\mbox{.}}{2015}]%
        {akoglu2015graph}
\bibfield{author}{\bibinfo{person}{Leman Akoglu}, \bibinfo{person}{Hanghang
  Tong}, {and} \bibinfo{person}{Danai Koutra}.}
  \bibinfo{year}{2015}\natexlab{}.
\newblock \showarticletitle{Graph based anomaly detection and description: a
  survey}.
\newblock \bibinfo{journal}{\emph{Data Mining and Knowledge Discovery}}
  \bibinfo{volume}{29}, \bibinfo{number}{3} (\bibinfo{year}{2015}),
  \bibinfo{pages}{626--688}.
\newblock


\bibitem[\protect\citeauthoryear{Broido and Clauset}{Broido and
  Clauset}{2019}]%
        {broido2019scale}
\bibfield{author}{\bibinfo{person}{Anna~D Broido} {and} \bibinfo{person}{Aaron
  Clauset}.} \bibinfo{year}{2019}\natexlab{}.
\newblock \showarticletitle{Scale-free networks are rare}.
\newblock \bibinfo{journal}{\emph{Nature Communications}} \bibinfo{volume}{10},
  \bibinfo{number}{1} (\bibinfo{year}{2019}), \bibinfo{pages}{1017}.
\newblock


\bibitem[\protect\citeauthoryear{Carlini and Wagner}{Carlini and
  Wagner}{2017}]%
        {carlini2017towards}
\bibfield{author}{\bibinfo{person}{Nicholas Carlini} {and}
  \bibinfo{person}{David Wagner}.} \bibinfo{year}{2017}\natexlab{}.
\newblock \showarticletitle{Towards evaluating the robustness of neural
  networks}. In \bibinfo{booktitle}{\emph{IEEE Symposium on Security and
  Privacy (SP)}}. \bibinfo{pages}{39--57}.
\newblock


\bibitem[\protect\citeauthoryear{Cavallari, Zheng, Cai, Chang, and
  Cambria}{Cavallari et~al\mbox{.}}{2017}]%
        {cavallari2017learning}
\bibfield{author}{\bibinfo{person}{Sandro Cavallari},
  \bibinfo{person}{Vincent~W Zheng}, \bibinfo{person}{Hongyun Cai},
  \bibinfo{person}{Kevin Chen-Chuan Chang}, {and} \bibinfo{person}{Erik
  Cambria}.} \bibinfo{year}{2017}\natexlab{}.
\newblock \showarticletitle{Learning community embedding with community
  detection and node embedding on graphs}. In \bibinfo{booktitle}{\emph{CIKM}}.
  \bibinfo{pages}{377--386}.
\newblock


\bibitem[\protect\citeauthoryear{Chen, Chen, Chen, Zhao, Yu, Xuan, and
  Yang}{Chen et~al\mbox{.}}{2018}]%
        {Chen2018GABasedQO}
\bibfield{author}{\bibinfo{person}{Jinyin Chen}, \bibinfo{person}{Lihong Chen},
  \bibinfo{person}{Yixian Chen}, \bibinfo{person}{Minghao Zhao},
  \bibinfo{person}{Shanqing Yu}, \bibinfo{person}{Qi Xuan}, {and}
  \bibinfo{person}{Xiaoniu Yang}.} \bibinfo{year}{2018}\natexlab{}.
\newblock \showarticletitle{GA-Based Q-Attack on Community Detection}.
\newblock \bibinfo{journal}{\emph{IEEE Transactions on Computational Social
  Systems}}  \bibinfo{volume}{6} (\bibinfo{year}{2018}),
  \bibinfo{pages}{491--503}.
\newblock


\bibitem[\protect\citeauthoryear{Chen, Li, and Bruna}{Chen
  et~al\mbox{.}}{2019}]%
        {chen2017supervised}
\bibfield{author}{\bibinfo{person}{Zhengdao Chen}, \bibinfo{person}{Xiang Li},
  {and} \bibinfo{person}{Joan Bruna}.} \bibinfo{year}{2019}\natexlab{}.
\newblock \showarticletitle{Supervised community detection with line graph
  neural networks}. In \bibinfo{booktitle}{\emph{ICLR}}.
\newblock


\bibitem[\protect\citeauthoryear{Dai, Li, Tian, Huang, Wang, Zhu, and Song}{Dai
  et~al\mbox{.}}{2018}]%
        {dai2018adversarial}
\bibfield{author}{\bibinfo{person}{Hanjun Dai}, \bibinfo{person}{Hui Li},
  \bibinfo{person}{Tian Tian}, \bibinfo{person}{Xin Huang},
  \bibinfo{person}{Lin Wang}, \bibinfo{person}{Jun Zhu}, {and}
  \bibinfo{person}{Le Song}.} \bibinfo{year}{2018}\natexlab{}.
\newblock \showarticletitle{Adversarial attack on graph structured data}. In
  \bibinfo{booktitle}{\emph{ICML}}. \bibinfo{pages}{1123--1132}.
\newblock


\bibitem[\protect\citeauthoryear{Fionda and Pirro}{Fionda and Pirro}{2017}]%
        {fionda2017community}
\bibfield{author}{\bibinfo{person}{Valeria Fionda} {and}
  \bibinfo{person}{Giuseppe Pirro}.} \bibinfo{year}{2017}\natexlab{}.
\newblock \showarticletitle{Community deception or: How to stop fearing
  community detection algorithms}.
\newblock \bibinfo{journal}{\emph{IEEE Transactions on Knowledge and Data
  Engineering}} \bibinfo{volume}{30}, \bibinfo{number}{4}
  (\bibinfo{year}{2017}), \bibinfo{pages}{660--673}.
\newblock


\bibitem[\protect\citeauthoryear{Goodfellow, Pouget-Abadie, Mirza, Xu,
  Warde-Farley, Ozair, Courville, and Bengio}{Goodfellow et~al\mbox{.}}{2014}]%
        {goodfellow2014generative}
\bibfield{author}{\bibinfo{person}{Ian Goodfellow}, \bibinfo{person}{Jean
  Pouget-Abadie}, \bibinfo{person}{Mehdi Mirza}, \bibinfo{person}{Bing Xu},
  \bibinfo{person}{David Warde-Farley}, \bibinfo{person}{Sherjil Ozair},
  \bibinfo{person}{Aaron Courville}, {and} \bibinfo{person}{Yoshua Bengio}.}
  \bibinfo{year}{2014}\natexlab{}.
\newblock \showarticletitle{Generative adversarial nets}. In
  \bibinfo{booktitle}{\emph{NIPS}}. \bibinfo{pages}{2672--2680}.
\newblock


\bibitem[\protect\citeauthoryear{Grover and Leskovec}{Grover and
  Leskovec}{2016}]%
        {grover2016node2vec}
\bibfield{author}{\bibinfo{person}{Aditya Grover} {and} \bibinfo{person}{Jure
  Leskovec}.} \bibinfo{year}{2016}\natexlab{}.
\newblock \showarticletitle{node2vec: Scalable feature learning for networks}.
  In \bibinfo{booktitle}{\emph{KDD}}. \bibinfo{pages}{855--864}.
\newblock


\bibitem[\protect\citeauthoryear{Grover, Zweig, and Ermon}{Grover
  et~al\mbox{.}}{2019}]%
        {grover2018graphite}
\bibfield{author}{\bibinfo{person}{Aditya Grover}, \bibinfo{person}{Aaron
  Zweig}, {and} \bibinfo{person}{Stefano Ermon}.}
  \bibinfo{year}{2019}\natexlab{}.
\newblock \showarticletitle{Graphite: Iterative generative modeling of graphs}.
  In \bibinfo{booktitle}{\emph{ICML}}. \bibinfo{pages}{2434--2444}.
\newblock


\bibitem[\protect\citeauthoryear{Kingma and Welling}{Kingma and
  Welling}{2014}]%
        {kingma2013auto}
\bibfield{author}{\bibinfo{person}{Diederik~P Kingma} {and}
  \bibinfo{person}{Max Welling}.} \bibinfo{year}{2014}\natexlab{}.
\newblock \showarticletitle{Auto-encoding variational bayes}. In
  \bibinfo{booktitle}{\emph{ICLR}}.
\newblock


\bibitem[\protect\citeauthoryear{Kipf and Welling}{Kipf and Welling}{2016}]%
        {kipf2016variational}
\bibfield{author}{\bibinfo{person}{Thomas~N Kipf} {and} \bibinfo{person}{Max
  Welling}.} \bibinfo{year}{2016}\natexlab{}.
\newblock \showarticletitle{Variational Graph Auto-Encoders}. In
  \bibinfo{booktitle}{\emph{NIPS Workshop on Bayesian Deep Learning}}.
\newblock


\bibitem[\protect\citeauthoryear{Kipf and Welling}{Kipf and Welling}{2017}]%
        {kipf2017semi}
\bibfield{author}{\bibinfo{person}{Thomas~N. Kipf} {and} \bibinfo{person}{Max
  Welling}.} \bibinfo{year}{2017}\natexlab{}.
\newblock \showarticletitle{Semi-Supervised Classification with Graph
  Convolutional Networks}. In \bibinfo{booktitle}{\emph{ICLR}}.
\newblock


\bibitem[\protect\citeauthoryear{Klicpera, Bojchevski, and
  G{\"u}nnemann}{Klicpera et~al\mbox{.}}{2019}]%
        {klicpera2018predict}
\bibfield{author}{\bibinfo{person}{Johannes Klicpera},
  \bibinfo{person}{Aleksandar Bojchevski}, {and} \bibinfo{person}{Stephan
  G{\"u}nnemann}.} \bibinfo{year}{2019}\natexlab{}.
\newblock \showarticletitle{Predict then propagate: Graph neural networks meet
  personalized pagerank}. In \bibinfo{booktitle}{\emph{ICLR}}.
\newblock


\bibitem[\protect\citeauthoryear{Konda and Tsitsiklis}{Konda and
  Tsitsiklis}{2000}]%
        {konda2000actor}
\bibfield{author}{\bibinfo{person}{Vijay~R Konda} {and} \bibinfo{person}{John~N
  Tsitsiklis}.} \bibinfo{year}{2000}\natexlab{}.
\newblock \showarticletitle{Actor-critic algorithms}. In
  \bibinfo{booktitle}{\emph{NIPS}}. \bibinfo{pages}{1008--1014}.
\newblock


\bibitem[\protect\citeauthoryear{Koutra, Parikh, Ramdas, and Xiang}{Koutra
  et~al\mbox{.}}{2011}]%
        {koutra2011algorithms}
\bibfield{author}{\bibinfo{person}{Danai Koutra}, \bibinfo{person}{Ankur
  Parikh}, \bibinfo{person}{Aaditya Ramdas}, {and} \bibinfo{person}{Jing
  Xiang}.} \bibinfo{year}{2011}\natexlab{}.
\newblock \showarticletitle{Algorithms for graph similarity and subgraph
  matching}.
\newblock


\bibitem[\protect\citeauthoryear{Kusner, Paige, and
  Hern{\'a}ndez-Lobato}{Kusner et~al\mbox{.}}{2017}]%
        {kusner2017grammar}
\bibfield{author}{\bibinfo{person}{Matt~J Kusner}, \bibinfo{person}{Brooks
  Paige}, {and} \bibinfo{person}{Jos{\'e}~Miguel Hern{\'a}ndez-Lobato}.}
  \bibinfo{year}{2017}\natexlab{}.
\newblock \showarticletitle{Grammar variational autoencoder}. In
  \bibinfo{booktitle}{\emph{ICML}}. \bibinfo{pages}{1945--1954}.
\newblock


\bibitem[\protect\citeauthoryear{Li, Rong, Cheng, Meng, Huang, and Huang}{Li
  et~al\mbox{.}}{2019}]%
        {jiawww19}
\bibfield{author}{\bibinfo{person}{Jia Li}, \bibinfo{person}{Yu Rong},
  \bibinfo{person}{Hong Cheng}, \bibinfo{person}{Helen Meng},
  \bibinfo{person}{Wenbing Huang}, {and} \bibinfo{person}{Junzhou Huang}.}
  \bibinfo{year}{2019}\natexlab{}.
\newblock \showarticletitle{Semi-Supervised Graph Classification: A
  Hierarchical Graph Perspective}. In \bibinfo{booktitle}{\emph{WWW}}.
  \bibinfo{pages}{972--982}.
\newblock


\bibitem[\protect\citeauthoryear{Liu, Allamanis, Brockschmidt, and Gaunt}{Liu
  et~al\mbox{.}}{2018}]%
        {liu2018constrained}
\bibfield{author}{\bibinfo{person}{Qi Liu}, \bibinfo{person}{Miltiadis
  Allamanis}, \bibinfo{person}{Marc Brockschmidt}, {and}
  \bibinfo{person}{Alexander Gaunt}.} \bibinfo{year}{2018}\natexlab{}.
\newblock \showarticletitle{Constrained graph variational autoencoders for
  molecule design}. In \bibinfo{booktitle}{\emph{NeurIPS}}.
  \bibinfo{pages}{7795--7804}.
\newblock


\bibitem[\protect\citeauthoryear{Nagaraja}{Nagaraja}{2010}]%
        {nagaraja2010impact}
\bibfield{author}{\bibinfo{person}{Shishir Nagaraja}.}
  \bibinfo{year}{2010}\natexlab{}.
\newblock \showarticletitle{The impact of unlinkability on adversarial
  community detection: effects and countermeasures}. In
  \bibinfo{booktitle}{\emph{International Symposium on Privacy Enhancing
  Technologies Symposium}}. \bibinfo{pages}{253--272}.
\newblock


\bibitem[\protect\citeauthoryear{Nazi, Hang, Goldie, Ravi, and Mirhoseini}{Nazi
  et~al\mbox{.}}{2019}]%
        {nazi2019gap}
\bibfield{author}{\bibinfo{person}{Azade Nazi}, \bibinfo{person}{Will Hang},
  \bibinfo{person}{Anna Goldie}, \bibinfo{person}{Sujith Ravi}, {and}
  \bibinfo{person}{Azalia Mirhoseini}.} \bibinfo{year}{2019}\natexlab{}.
\newblock \showarticletitle{GAP: Generalizable Approximate Graph Partitioning
  Framework}. In \bibinfo{booktitle}{\emph{ICLR workshop}}.
\newblock


\bibitem[\protect\citeauthoryear{Newman}{Newman}{2006}]%
        {newman2006modularity}
\bibfield{author}{\bibinfo{person}{Mark~EJ Newman}.}
  \bibinfo{year}{2006}\natexlab{}.
\newblock \showarticletitle{Modularity and community structure in networks}.
\newblock \bibinfo{journal}{\emph{Proceedings of the National Academy of
  Sciences}} \bibinfo{volume}{103}, \bibinfo{number}{23}
  (\bibinfo{year}{2006}), \bibinfo{pages}{8577--8582}.
\newblock


\bibitem[\protect\citeauthoryear{Page, Brin, Motwani, and Winograd}{Page
  et~al\mbox{.}}{1999}]%
        {ilprints422}
\bibfield{author}{\bibinfo{person}{Lawrence Page}, \bibinfo{person}{Sergey
  Brin}, \bibinfo{person}{Rajeev Motwani}, {and} \bibinfo{person}{Terry
  Winograd}.} \bibinfo{year}{1999}\natexlab{}.
\newblock \bibinfo{booktitle}{\emph{The PageRank Citation Ranking: Bringing
  Order to the Web.}}
\newblock \bibinfo{type}{Technical Report} 1999-66.
\newblock
\newblock
\shownote{Previous number = SIDL-WP-1999-0120.}


\bibitem[\protect\citeauthoryear{Perozzi, Al-Rfou, and Skiena}{Perozzi
  et~al\mbox{.}}{2014}]%
        {perozzi2014deepwalk}
\bibfield{author}{\bibinfo{person}{Bryan Perozzi}, \bibinfo{person}{Rami
  Al-Rfou}, {and} \bibinfo{person}{Steven Skiena}.}
  \bibinfo{year}{2014}\natexlab{}.
\newblock \showarticletitle{Deepwalk: Online learning of social
  representations}. In \bibinfo{booktitle}{\emph{KDD}}.
  \bibinfo{pages}{701--710}.
\newblock


\bibitem[\protect\citeauthoryear{Rosvall and Bergstrom}{Rosvall and
  Bergstrom}{2008}]%
        {rosvall2008maps}
\bibfield{author}{\bibinfo{person}{Martin Rosvall} {and}
  \bibinfo{person}{Carl~T Bergstrom}.} \bibinfo{year}{2008}\natexlab{}.
\newblock \showarticletitle{Maps of random walks on complex networks reveal
  community structure}.
\newblock \bibinfo{journal}{\emph{Proceedings of the National Academy of
  Sciences}} \bibinfo{volume}{105}, \bibinfo{number}{4} (\bibinfo{year}{2008}),
  \bibinfo{pages}{1118--1123}.
\newblock


\bibitem[\protect\citeauthoryear{Samanta, Abir, Jana, Chattaraj, Ganguly, and
  Rodriguez}{Samanta et~al\mbox{.}}{2019}]%
        {samanta2019nevae}
\bibfield{author}{\bibinfo{person}{Bidisha Samanta}, \bibinfo{person}{DE Abir},
  \bibinfo{person}{Gourhari Jana}, \bibinfo{person}{Pratim~Kumar Chattaraj},
  \bibinfo{person}{Niloy Ganguly}, {and} \bibinfo{person}{Manuel~Gomez
  Rodriguez}.} \bibinfo{year}{2019}\natexlab{}.
\newblock \showarticletitle{Nevae: A deep generative model for molecular
  graphs}. In \bibinfo{booktitle}{\emph{AAAI}}. \bibinfo{pages}{1110--1117}.
\newblock


\bibitem[\protect\citeauthoryear{Shaham, Stanton, Li, Nadler, Basri, and
  Kluger}{Shaham et~al\mbox{.}}{2018}]%
        {shaham2018spectralnet}
\bibfield{author}{\bibinfo{person}{Uri Shaham}, \bibinfo{person}{Kelly
  Stanton}, \bibinfo{person}{Henry Li}, \bibinfo{person}{Boaz Nadler},
  \bibinfo{person}{Ronen Basri}, {and} \bibinfo{person}{Yuval Kluger}.}
  \bibinfo{year}{2018}\natexlab{}.
\newblock \showarticletitle{Spectralnet: Spectral clustering using deep neural
  networks}. In \bibinfo{booktitle}{\emph{ICLR}}.
\newblock


\bibitem[\protect\citeauthoryear{Shi and Malik}{Shi and Malik}{2000}]%
        {shi2000normalized}
\bibfield{author}{\bibinfo{person}{Jianbo Shi} {and} \bibinfo{person}{Jitendra
  Malik}.} \bibinfo{year}{2000}\natexlab{}.
\newblock \showarticletitle{Normalized cuts and image segmentation}.
\newblock \bibinfo{journal}{\emph{IEEE Transactions on Pattern Analysis and
  Machine Intelligence}} \bibinfo{volume}{22}, \bibinfo{number}{8}
  (\bibinfo{year}{2000}), \bibinfo{pages}{888--905}.
\newblock


\bibitem[\protect\citeauthoryear{Sun, Wang, Yu, and Li}{Sun
  et~al\mbox{.}}{2018}]%
        {sun2018adversarial}
\bibfield{author}{\bibinfo{person}{Lichao Sun}, \bibinfo{person}{Ji Wang},
  \bibinfo{person}{Philip~S Yu}, {and} \bibinfo{person}{Bo Li}.}
  \bibinfo{year}{2018}\natexlab{}.
\newblock \showarticletitle{Adversarial attack and defense on graph data: A
  survey}.
\newblock \bibinfo{journal}{\emph{arXiv preprint arXiv:1812.10528}}
  (\bibinfo{year}{2018}).
\newblock


\bibitem[\protect\citeauthoryear{Tang, Djelouah, Perazzi, Boykov, and
  Schroers}{Tang et~al\mbox{.}}{2018}]%
        {tang2018normalized}
\bibfield{author}{\bibinfo{person}{Meng Tang}, \bibinfo{person}{Abdelaziz
  Djelouah}, \bibinfo{person}{Federico Perazzi}, \bibinfo{person}{Yuri Boykov},
  {and} \bibinfo{person}{Christopher Schroers}.}
  \bibinfo{year}{2018}\natexlab{}.
\newblock \showarticletitle{Normalized cut loss for weakly-supervised cnn
  segmentation}. In \bibinfo{booktitle}{\emph{CVPR}}.
  \bibinfo{pages}{1818--1827}.
\newblock


\bibitem[\protect\citeauthoryear{Wang, Wang, Yu, and Zhang}{Wang
  et~al\mbox{.}}{2015}]%
        {wang2015community}
\bibfield{author}{\bibinfo{person}{Meng Wang}, \bibinfo{person}{Chaokun Wang},
  \bibinfo{person}{Jeffrey~Xu Yu}, {and} \bibinfo{person}{Jun Zhang}.}
  \bibinfo{year}{2015}\natexlab{}.
\newblock \showarticletitle{Community detection in social networks: an in-depth
  benchmarking study with a procedure-oriented framework}.
\newblock \bibinfo{journal}{\emph{Proceedings of the VLDB Endowment}}
  \bibinfo{volume}{8}, \bibinfo{number}{10} (\bibinfo{year}{2015}),
  \bibinfo{pages}{998--1009}.
\newblock


\bibitem[\protect\citeauthoryear{Waniek, Michalak, Wooldridge, and
  Rahwan}{Waniek et~al\mbox{.}}{2018}]%
        {waniek2018hiding}
\bibfield{author}{\bibinfo{person}{Marcin Waniek}, \bibinfo{person}{Tomasz~P
  Michalak}, \bibinfo{person}{Michael~J Wooldridge}, {and}
  \bibinfo{person}{Talal Rahwan}.} \bibinfo{year}{2018}\natexlab{}.
\newblock \showarticletitle{Hiding individuals and communities in a social
  network}.
\newblock \bibinfo{journal}{\emph{Nature Human Behaviour}} \bibinfo{volume}{2},
  \bibinfo{number}{2} (\bibinfo{year}{2018}), \bibinfo{pages}{139}.
\newblock


\bibitem[\protect\citeauthoryear{Wu, Pan, Chen, Long, Zhang, and Yu}{Wu
  et~al\mbox{.}}{2019}]%
        {wu2019comprehensive}
\bibfield{author}{\bibinfo{person}{Zonghan Wu}, \bibinfo{person}{Shirui Pan},
  \bibinfo{person}{Fengwen Chen}, \bibinfo{person}{Guodong Long},
  \bibinfo{person}{Chengqi Zhang}, {and} \bibinfo{person}{Philip~S Yu}.}
  \bibinfo{year}{2019}\natexlab{}.
\newblock \showarticletitle{A comprehensive survey on graph neural networks}.
\newblock \bibinfo{journal}{\emph{arXiv preprint arXiv:1901.00596}}
  (\bibinfo{year}{2019}).
\newblock


\bibitem[\protect\citeauthoryear{Zhou, Cheng, and Yu}{Zhou
  et~al\mbox{.}}{2009}]%
        {zhou2009graph}
\bibfield{author}{\bibinfo{person}{Yang Zhou}, \bibinfo{person}{Hong Cheng},
  {and} \bibinfo{person}{Jeffrey~Xu Yu}.} \bibinfo{year}{2009}\natexlab{}.
\newblock \showarticletitle{Graph clustering based on structural/attribute
  similarities}.
\newblock \bibinfo{journal}{\emph{Proceedings of the VLDB Endowment}}
  \bibinfo{volume}{2}, \bibinfo{number}{1} (\bibinfo{year}{2009}),
  \bibinfo{pages}{718--729}.
\newblock


\bibitem[\protect\citeauthoryear{Z{\"u}gner, Akbarnejad, and
  G{\"u}nnemann}{Z{\"u}gner et~al\mbox{.}}{2018}]%
        {zugner2018adversarial}
\bibfield{author}{\bibinfo{person}{Daniel Z{\"u}gner}, \bibinfo{person}{Amir
  Akbarnejad}, {and} \bibinfo{person}{Stephan G{\"u}nnemann}.}
  \bibinfo{year}{2018}\natexlab{}.
\newblock \showarticletitle{Adversarial Attacks on Neural Networks for Graph
  Data}. In \bibinfo{booktitle}{\emph{KDD}}. \bibinfo{pages}{2847--2856}.
\newblock


\end{thebibliography}

\end{document}